\documentclass[twocolumn,amsmath,floatfix,amssymb,aps,prb,showpacs,letter,superscriptaddress]{revtex4-1}
\usepackage{graphicx,natbib}
\usepackage{color}

\begin{document}

\title{Antiferromagnetic Spin Ice Correlations at ($\frac{1}{2}$,$\frac{1}{2}$,$\frac{1}{2}$) in the Ground State of the Pyrochlore Magnet Tb$_{2}$Ti$_{2}$O$_{7}$}

\author{K. Fritsch}
\affiliation{Department of Physics and Astronomy, McMaster University, Hamilton, Ontario, L8S 4M1, Canada}
\author{K. A. Ross}
\affiliation{Department of Physics and Astronomy, McMaster University, Hamilton, Ontario, L8S 4M1, Canada}
\affiliation{Institute for Quantum Matter and Department of Physics and Astronomy, Johns Hopkins University, Baltimore, Maryland 21218, USA}
\affiliation{NIST Center for Neutron Research, NIST, Gaithersburg, Maryland 20899-8102, USA}
\author{Y. Qiu}
\affiliation{NIST Center for Neutron Research, NIST, Gaithersburg, Maryland 20899-8102, USA}
\affiliation{Department of Materials Science and Engineering, University of Maryland, College Park, Maryland 20742, USA}
\author{J. R. D. Copley}
\affiliation{NIST Center for Neutron Research, NIST, Gaithersburg, Maryland 20899-8102, USA}
\author{T.~Guidi}
\affiliation{ISIS Pulsed Neutron and Muon Facility, Rutherford Appleton Laboratory, Chilton, Didcot OX11 0QX, UK}
\author{R. I. Bewley}
\affiliation{ISIS Pulsed Neutron and Muon Facility, Rutherford Appleton Laboratory, Chilton, Didcot OX11 0QX, UK}
\author{H. A. Dabkowska}
\affiliation{Brockhouse Institute for Materials Research, Hamilton, Ontario, L8S 4M1, Canada}
\author{B. D. Gaulin}
\affiliation{Department of Physics and Astronomy, McMaster University, Hamilton, Ontario, L8S 4M1, Canada}
\affiliation{Brockhouse Institute for Materials Research, Hamilton, Ontario, L8S 4M1, Canada}
\affiliation{Canadian Institute for Advanced Research, 180 Dundas St.\ W., Toronto, Ontario, M5G 1Z8, Canada}

\begin{abstract}
We present high-resolution single crystal time-of-flight neutron scattering measurements on the candidate quantum spin liquid pyrochlore Tb$_{2}$Ti$_{2}$O$_{7}$ at low temperature and in a magnetic field. At $\sim$70 mK and in zero field, Tb$_{2}$Ti$_{2}$O$_{7}$ reveals diffuse magnetic elastic scattering at ($\frac{1}{2}$,$\frac{1}{2}$,$\frac{1}{2}$) positions in reciprocal space, consistent with short-range correlated regions based on a two-in, two-out spin ice configuration on a doubled conventional unit cell. This elastic scattering is separated from very low-energy magnetic inelastic scattering by an energy gap of $\sim$ 0.06-0.08 meV. The elastic signal disappears under the application of small magnetic fields and upon elevating temperature. Pinch-point-like elastic diffuse scattering is observed near (1,1,1) and (0,0,2) in zero field at $\sim$70 mK, in agreement with Fennell {\it et al.} \cite{Fennell2012}, supporting the quantum spin ice interpretation of Tb$_2$Ti$_2$O$_7$.

\end{abstract}
\maketitle

\section{Introduction}

Geometrically frustrated magnets have been intensely studied over the course of the last decade due to the rich variety of unconventional ground states they display \cite{frustratedspinsystems}. In particular, the magnetic rare-earth titanate pyrochlore oxides with the chemical formula R$_2$Ti$_2$O$_7$ have received considerable attention \cite{JasonRMP}. In this structure, trivalent R$^{3+}$ rare-earth metal ions occupy a network of corner-sharing tetrahedra known as the pyrochlore lattice, which is the prototypical example of geometric frustration in three dimensions. The interplay between exchange and dipolar interactions with crystal-field (CF) induced anisotropy on the underlying pyrochlore lattice leads to a variety of exotic phenomena. These include ground state selection by order-by-disorder \cite{Savary,ZhitomirskyPRL}, as well as classical \cite{BramwellGingras,Gingrasin2,Henley,CastelnovoRev} and quantum spin ice \cite{OnodaGenericQSI2012,OnodaTanaka,OnodaPRB2011,KatePRX,Applegate2012,ChangOnoda2012} physics.

Tb$_{2}$Ti$_{2}$O$_{7}$ has attracted much interest as a potential experimental realization of a quantum spin liquid \cite{BalentsSL}, based on its lack of long range order (LRO) down to at least 50 mK \cite{Jason1999,Jason2003} despite a Curie Weiss temperature of $\sim$-14 K \cite{GingrasPRB2000}. The Tb$^{3+}$ ions in Tb$_{2}$Ti$_{2}$O$_{7}$ have an Ising CF doublet ground state that leads to $<$111$>$ easy-axis anisotropy as in the canonical spin ice materials Ho$_{2}$Ti$_{2}$O$_{7}$ and Dy$_{2}$Ti$_{2}$O$_{7}$. In contrast to these spin ices however, the net exchange in Tb$_{2}$Ti$_{2}$O$_{7}$ is antiferromagnetic (AF). Since the combination of Ising-like $<$111$>$ anisotropy and AF interactions on the pyrochlore lattice is unfrustrated \cite{Bramwell}, Tb$_{2}$Ti$_{2}$O$_{7}$ is naively expected to display a unique LRO ground state.

Recently, two theoretical pictures have sought to explain its enigmatic ground state. One of these proposed a ``quantum spin ice'' (QSI) scenario for the disordered ground state in Tb$_{2}$Ti$_{2}$O$_{7}$ \cite{MolavianQSI2007,MolavianQSI2009}. This scenario invokes virtual quantum excitations between the CF ground state doublet and the first excited doublet ($\Delta\sim$15 K) that reposition Tb$_{2}$Ti$_{2}$O$_{7}$ into the spin ice regime. Jahn-Teller physics and a non-magnetic singlet ground state have also been proposed and debated \cite{BonvilleSinglet2011,BruceCF} as the cause underlying its failure to order.

\begin{figure*}
\includegraphics[width=\textwidth]{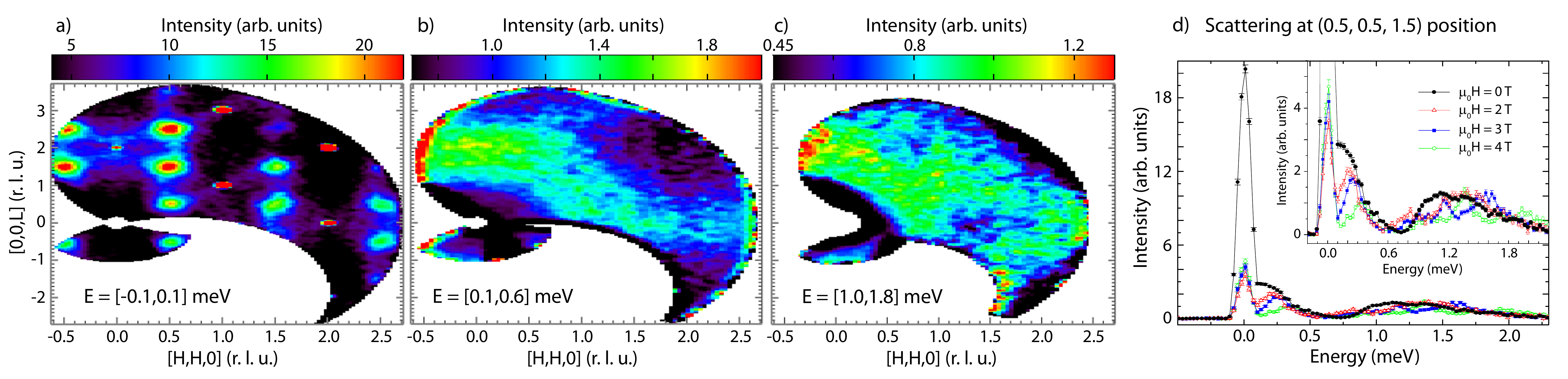}%
\caption{Neutron scattering data within the (H,H,L) plane of Tb$_{2}$Ti$_{2}$O$_{7}$ at T = 70 mK are shown for a) -0.1 meV $<$ E $<$ 0.1 meV, b) 0.1 meV $<$ E $<$ 0.6 meV and c) 1.0 $<$ E $<$ 1.8 meV. Panel d) shows a plot of intensity vs energy transfer at the ($\frac{1}{2}$,$\frac{1}{2}$,$\frac{3}{2}$) position for fields $\mu_0$H=0,2,3, and 4 T. The integration  range is H=[0.2,0.8] r.l.u., L=[1.2,1.8] r.l.u.. The inset expands the intensity scale. All data shown were corrected for detector efficiency, and an empty can background was subtracted. The error bars are $\pm$1$\sigma$.}%
\label{figure1}%
\end{figure*}

In this paper, we report new neutron scattering measurements on single crystalline Tb$_{2}$Ti$_{2}$O$_{7}$ that elucidate the nature of its ground state. Our measurements were performed in zero field and in magnetic fields applied along two high-symmetry directions in the pyrochlore lattice. The zero field measurements reveal short-range AF spin ice correlations extending over roughly two conventional unit cells, characterized by elastic diffuse scattering around ($\frac{1}{2}$,$\frac{1}{2}$,$\frac{1}{2}$) positions in reciprocal space. The previously reported checkerboard-pattern of diffuse scattering \cite{Jason2001,Kirrily2006} is now resolved as low-energy inelastic scattering for 0.06 meV $<$ E $<$ 0.6 meV, while the higher-energy inelastic scattering regime for 0.8 meV $<$ E $<$ 2.0 meV is dominated by CF excitations, in agreement with earlier studies \cite{GingrasPRB2000,MirebeauPRB2007}. A model of the elastic scattering intensity involving 128 spins in a doubled conventional unit cell is presented. A fit of this model to the observed peak intensities suggests that the Tb$^{3+}$ moments form a short ranged AF ordered spin ice configuration with spins tilted $\sim$12$^\circ$ from their local $<$111$>$ axes. Finally, pinch-point like diffuse elastic scattering is observed near (0,0,2) in zero field. These results suggest that Tb$_{2}$Ti$_{2}$O$_{7}$ may well form a long range ordered (LRO) equilibrium state based on this spin ice-derived structure at lower temperatures, although this might require more pristine samples than are currently available.

\section{Experimental Details}
The single crystal sample of Tb$_{2}$Ti$_{2}$O$_{7}$ used for both neutron scattering measurements was grown using the optical floating zone technique at McMaster University \cite{HannaOFZ,Gardnercrystalgrowth}. It is the same single crystal used in an earlier study by Rule {\it et al.} \cite{Kirrily2006}. Time-of-flight neutron scattering measurements were performed using the disk-chopper spectrometer DCS \cite{CopleyDCS} at the NIST Center for Neutron Research and the LET spectrometer \cite{LET} at the ISIS Spallation Neutron Source. For the DCS measurements, incident neutrons of energy E$_i$=3.27 meV were employed, giving an energy resolution of 0.1 meV. The sample was carefully aligned with the [1-10] direction vertical to within 0.5$^\circ$, such that the [H,H,L] plane was coincident with the horizontal scattering plane. In the LET experiment, high resolution measurements using E$_i$=2.32 meV gave an energy resolution of 0.02 meV. At LET, the sample was aligned with the [11-1] direction vertical, placing the plane spanned by [H,H,2H] and [H,-H,0] within the horizontal scattering plane. Both DCS and LET experiments achieved a base temperature of $\sim$70 mK and maximum magnetic fields of $\mu_0$H = 10 T and $\mu_0$H = 7 T, respectively.

\section{Results and Discussion}

Figures 1(a)-1(c) show reciprocal lattice maps in the [H,H,L] plane in zero field, taken at T$\sim$ 70 mK on DCS. The data were corrected for detector efficiency, and to eliminate scattering from the sample environment. Figure 1(a) shows data in the elastic channel, integrating over -0.1 meV $<$ E $<$ 0.1 meV. These data reveal strong diffuse elastic scattering at ($\frac{1}{2}$,$\frac{1}{2}$,$\frac{1}{2}$)-type positions in reciprocal space. Cuts through the diffuse elastic scattering at ($\frac{3}{2}$,$\frac{3}{2}$,$\frac{3}{2}$) are shown in Fig. 2, for each of the [H,H,0], [H,H,H] and [0,0,L] directions, along with fits of this scattering to a standard Ornstein-Zernike (Lorentzian) form.  As can be seen from Fig. 2, the diffuse elastic scattering is isotropic in {\bf Q}, and characterized by a correlation length of $\xi=8.0\pm0.6$ \AA. This corresponds to a correlated static magnetic region of diameter 2$\xi$ = 16 \AA\ or roughly two conventional unit cells.  The elastic scattering we observe indicates spins which are static on the time scale of our energy resolution, $\sim$10$^{-10}$ s. 
\begin{figure}[h]
\includegraphics[width=8cm]{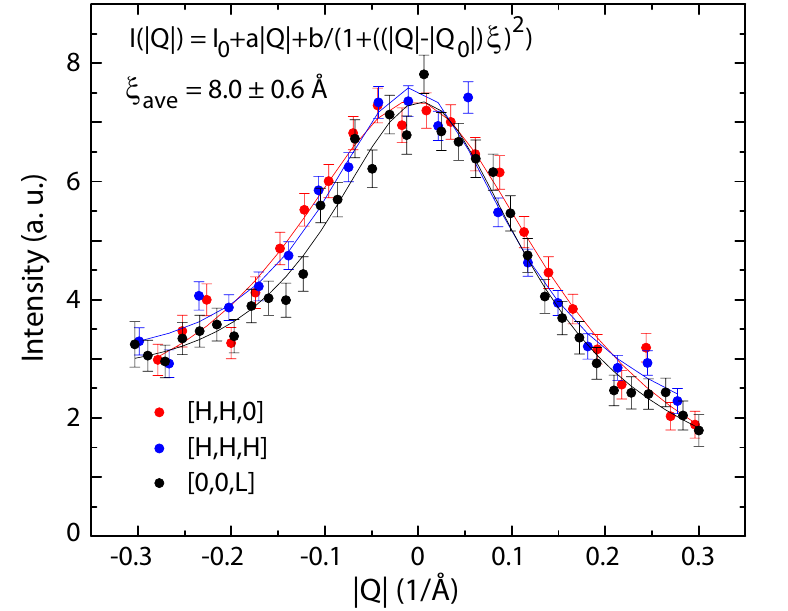}%
\caption{Cuts of the elastic scattering data in Fig. 1a) through the ($\frac{3}{2}$,$\frac{3}{2}$,$\frac{3}{2}$) position along each of the [H,H,0], [H,H,H], and [0,0,L] directions are shown, along with fits to an Ornstein-Zernike form for the diffuse lineshape. These data and associated fits show the short-range ordered, elastic antiferromagnetic Bragg features to be isotropic in {\bf Q}, and characterized by a correlation length of $\sim$8 \AA.}
\end{figure}

Figure 1(b) shows the low-lying inelastic scattering for 0.1 meV $<$ E $<$ 0.6 meV. Within our energy resolution, this scattering appears to be nearly quasielastic and forms the distinct checkerboard pattern observed previously \cite{Jason2001} with highest intensities near (0,0,2) and (2,2,0) in the [H,H,L] plane. Interestingly, this feature has boundaries formed by the ($\frac{1}{2}$,$\frac{1}{2}$,$\frac{1}{2}$) diffuse elastic scattering. The higher energy inelastic scattering for 1.0 meV $<$ E $<$ 1.8 meV is shown in Fig. 1(c). This intensity results from excitations to the lowest excited CF doublet states at $\Delta\sim$1.2 meV. Figure 1(d) shows the intensity around the ($\frac{1}{2}$,$\frac{1}{2}$,$\frac{3}{2}$) position as a function of energy for selected magnetic fields. The zero field data nicely illustrate the three distinct energy regimes that are mapped in reciprocal space in Figs. 1(a)-1(c). Under the application of a 2 T field, the zero field elastic scattering at the ($\frac{1}{2}$,$\frac{1}{2}$,$\frac{3}{2}$) and related positions drops to one-seventh of the initial intensity, as the elastic magnetic scattering has been eliminated, leaving only nuclear incoherent scattering at the ($\frac{1}{2}$,$\frac{1}{2}$,$\frac{3}{2}$) position. The low-lying inelastic scattering displays a clear peak at $\sim$0.2 meV at 2 T.  This feature moves to higher energies and fades in intensity in larger magnetic fields. The broad, higher energy inelastic scattering from the CF excitations also splits in the presence of a [1-10] magnetic field, forming spin wave bands which were previously reported \cite{Kirrily2006}. 

\begin{figure}[h]
\includegraphics[width=\columnwidth]{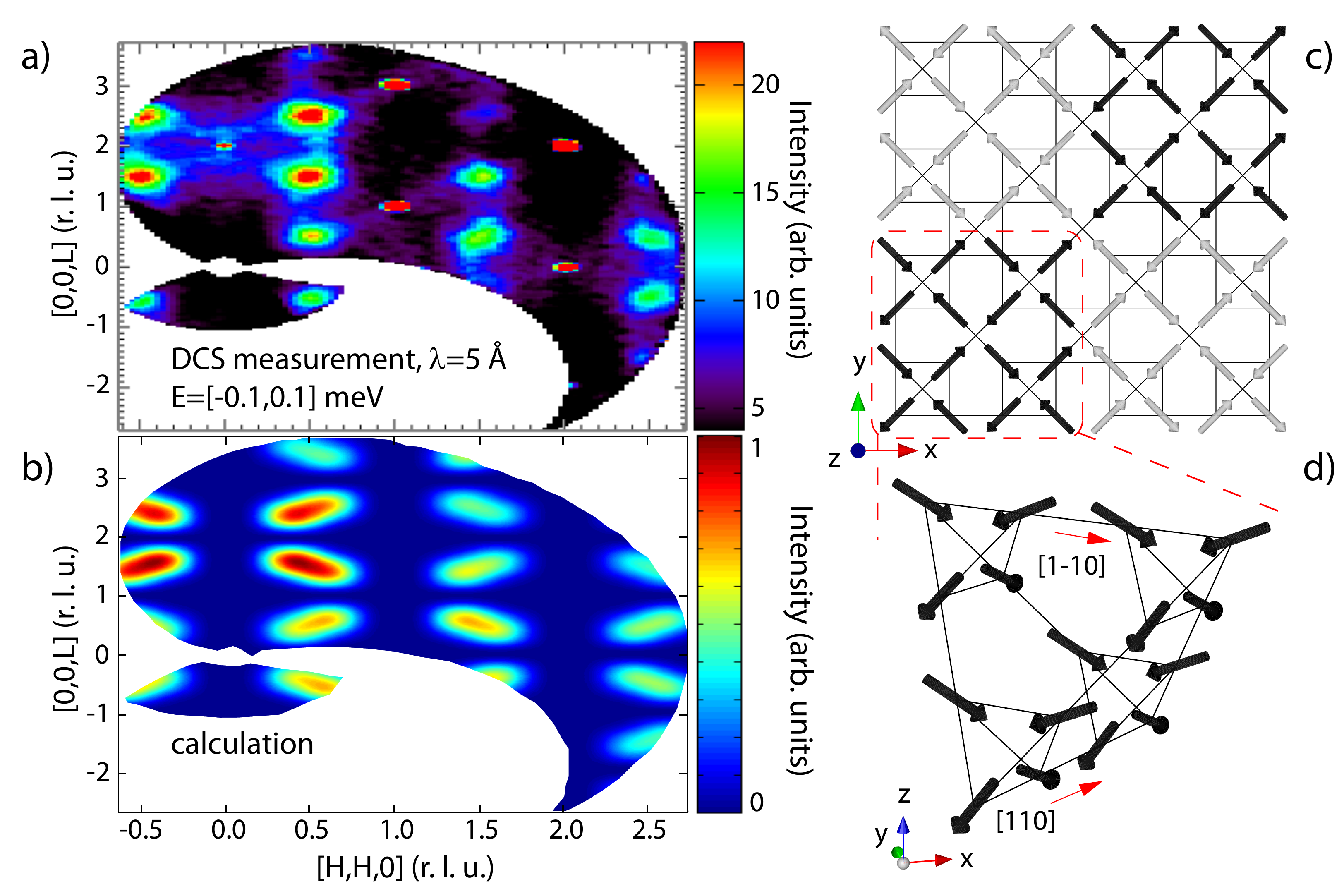}%
\caption{Comparison of the measured elastic diffuse scattering in the (H,H,L) plane of Tb$_{2}$Ti$_{2}$O$_{7}$ at 70 mK and $\mu_0$H = 0 (panel a)) to the calculated S({\bf{Q}}) discussed in the text (panel b)). The intensity scale in the calculation is normalized to the largest intensity at ($\frac{1}{2}$,$\frac{1}{2}$,$\frac{3}{2}$). The spin arrangement between neighboring conventional unit cells is shown as a projection onto the $xy$ plane in panel c). The ordered two-in, two-out spin configuration in a single unit cell is shown in panel d), and this pattern is reversed in the neighboring cells (black vs grey arrows in panel c)) to form ``$<\frac{1}{2},\frac{1}{2},\frac{1}{2}>$ ordered spin ice''. The spins are tilted from their local $<$111$>$ axes by $\sim$12$^\circ$.}%
\label{figure3}%
\end{figure}

The new elastic diffuse scattering at ($\frac{1}{2}$,$\frac{1}{2}$,$\frac{1}{2}$) and related wavevectors can be well described by a model based on a ``two-in, two-out'' spin ice arrangement that extends over two conventional unit cells in all three dimensions, leading to a supercell composed of eight conventional unit cells. This model takes into account a) the observed phase shift of $\pi$ between neighboring unit cells giving rise to scattering at ($\frac{1}{2}$,$\frac{1}{2}$,$\frac{1}{2}$) positions and b) the observed width of the diffuse scattering. The spin arrangement over two neighboring cells in the $xy$ plane is shown in Fig. 3 c). Note that the black spins are reversed in their neighboring cell (grey spins) along the $x$ or $y$ directions.

We calculate the neutron scattering intensity $S(\bf{Q}$) that is proportional to the square of the transverse component of the magnetic structure factor $|M_\perp({\bf{Q}})|^2=|{\bf{Q}} \!\times\!{\bf{M}}({\bf{Q}})\!\times\!{\bf{Q}}|^2$ with ${\bf{M}}({\bf{Q}})=\sum_jf_j({\bf{Q}}).{\bf{m}}_j.e^{i{\bf{Q}}\cdot{\bf{r}}_j}$, where $j$ runs over the 128 spins in the supercell. S(${\bf{Q}}$) is then averaged over all three spin ice domains. This calculation includes the magnetic form factor $f_j({\bf{Q}})$ for Tb$^{3+}$ and allows for direct comparison with experimental data \cite{formfactor}. In order to determine the deviation of the local structure from an ideal two-in, two-out scenario, we fit the calculated intensity to that at the nine ($\frac{1}{2}$,$\frac{1}{2}$,$\frac{1}{2}$)-related positions observed in the experiment, letting the canting angles from $<$111$>$ vary. The best agreement with the experimental data (Fig. 3 a)) is shown in Fig. 3 b). The calculated magnetic structure factor reproduces the net intensities of the diffuse ($\frac{1}{2}$,$\frac{1}{2}$,$\frac{1}{2}$)-like peaks very well. The elongation of the calculated diffuse scattering is a result of the finite size of the assumed spin arrangement (128 spins) for which we calculated the Fourier transform. The best-fit spin configuration has all spins canted by $\sim$12$^\circ$ from their local $<$111$>$ axis with a reduction of the magnetic moment along the local $z$ axis. Our fitting results are not very sensitive to the canting of the other two spin components. We show the resulting spin arrangement in Fig. 3 d) for one conventional unit cell. Tb$_{2}$Ti$_{2}$O$_{7}$ at 70mK is therefore composed of short range ordered (SRO) domains of an AF $<\frac{1}{2},\frac{1}{2},\frac{1}{2}>$ ordered spin ice.

The SRO AF spin-ice state of Tb$_{2}$Ti$_{2}$O$_{7}$ at $\sim$70 mK and zero field shows Tb$_{2}$Ti$_{2}$O$_{7}$ to have a strong tendency to order. An actual LRO state may require lower temperatures and equilibrium conditions.  Sensitivity to weak disorder, as is known to occur in single crystals of QSI Yb$_{2}$Ti$_{2}$O$_{7}$ \cite{KateYbTiOstructure}, may also be relevant. We note that the spin canting angle from the local $<$111$>$ axis is similar to the $\sim13^\circ$ spin cant observed in the ferromagnetic ``ordered spin ice'' phase in Tb$_{2}$Sn$_{2}$O$_{7}$ \cite{MirebeauPRL2005TSO,PetitTSOPRB2012}. We further note that a $\bf{q}$=($\frac{1}{2}$,$\frac{1}{2}$,$\frac{1}{2}$) ordering wavevector is known to be selected by the combination of isotropic near neighbor exchange and dipolar interactions, as occurs in Gd$_{2}$Ti$_{2}$O$_{7}$ \cite{ChampionGTO2001}.

\begin{figure}[h]
\includegraphics[width=8cm]{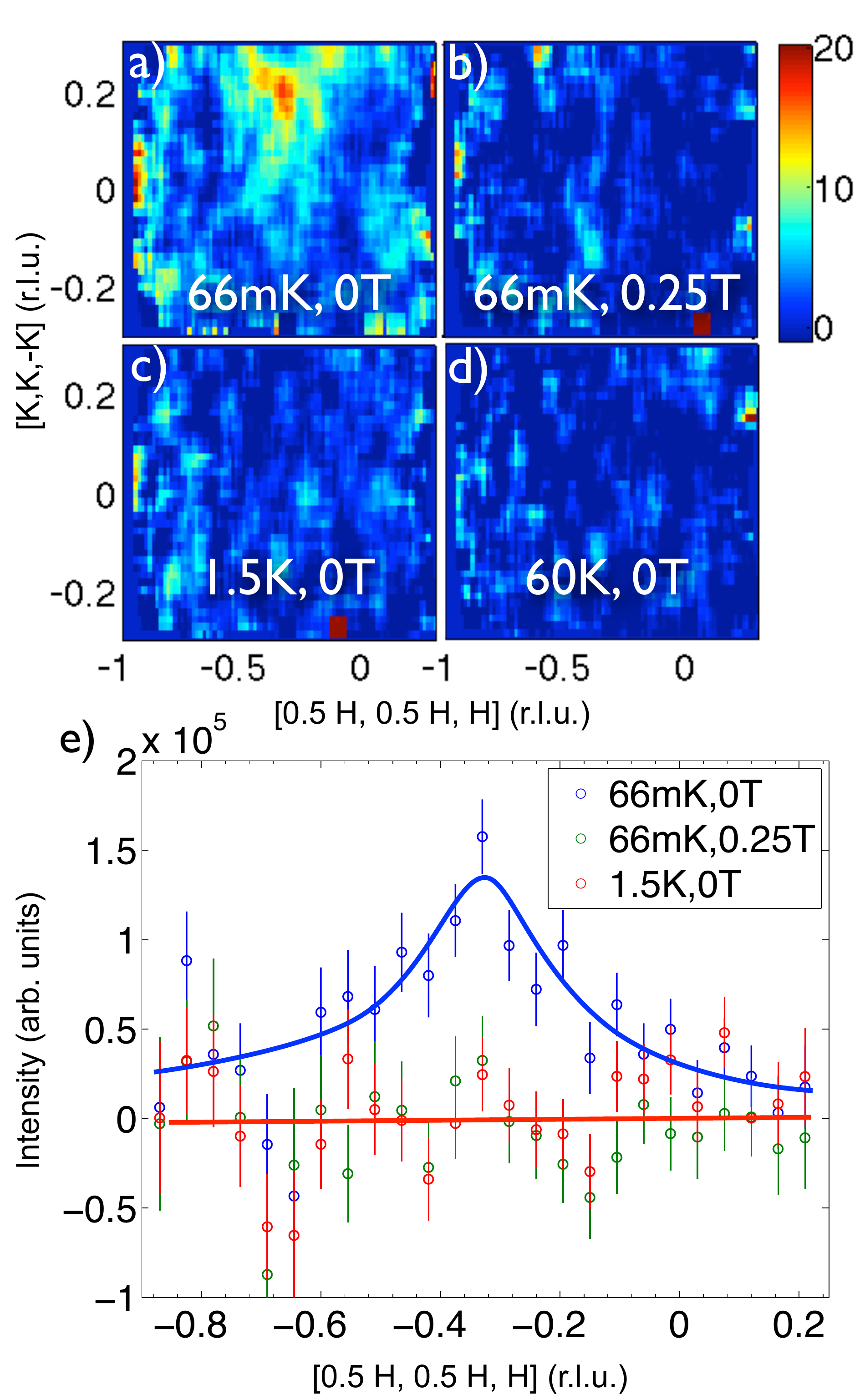}%
\caption{Elastic scattering data from LET, with the T=66 mK, $\mu_0$H=7 T data set used as a background, shown in the [K,K,-K] vs. [$\frac{1}{2}$H,$\frac{1}{2}$H,H] plane at $\mu_0$H=0, integrated from E=[-0.02, 0.02] meV, and [H,-H,0] = [-0.6,-0.4] r.l.u.. In the scattering geometry used in the LET experiment, the [K,K,-K] direction is parallel to the vertical (field) direction. a) The diffuse scattering at 66mK, 0T shown here corresponds to (-$\frac{1}{2}$,$\frac{1}{2}$,-$\frac{1}{2}$), which arises at [$\frac{1}{2}$H,$\frac{1}{2}$H,H] = -0.33 r.l.u., [H,-H,0] = -0.5 r.l.u., and [K,K,-K] = 0.166 r.l.u.. This diffuse elastic scattering disappears upon either application of a $\mu_0$H=0.25 T field at T=66 mK b), or warming to 1.5K (below $\Theta_{\textnormal{CW}}$),  c), or 60 K (above $\Theta_{\textnormal{CW}}$), d). Panel e) shows the field and temperature dependence of the diffuse elastic scattering at (-$\frac{1}{2}$,$\frac{1}{2}$,-$\frac{1}{2}$) as shown in panel a), integrating over [$\frac{1}{2}$H,$\frac{1}{2}$H,H] = [0,0.3] r.l.u..} 
\label{figure4}%
\end{figure}

Figures 4 and 5 show the field and temperature dependence of the scattering near the (-$\frac{1}{2}$,$\frac{1}{2}$,-$\frac{1}{2}$) position as measured on LET with an incident neutron energy of 2.32 meV. A high magnetic field, low temperature data set at T = 66 mK and $\mu_0$H=7 T has been subtracted from all data sets in Figures 4 and 5. As is shown in these figures, even a small magnetic field removes the elastic diffuse scattering at ($\frac{1}{2}$,$\frac{1}{2}$,$\frac{1}{2}$) wavevectors, and a large magnetic field such as $\mu_0$H=7 T, pushes all inelastic magnetic scattering to higher energies, leaving our energy range of interest ($\le$0.6 meV) empty.  Consequently our T=66 mK and $\mu_0$H=7 T data set serves as an excellent background for all the magnetic scattering of interest here.  

Figure 4 shows elastic scattering, binned between -0.02 meV $<$ E $<$ 0.02 meV for four different conditions of temperature and field, only one of which displays elastic Bragg-like scattering at ($\frac{1}{2}$,$\frac{1}{2}$,$\frac{1}{2}$). Figure 4 a) shows elastic magnetic Bragg-like scattering at ($\frac{1}{2}$,$\frac{1}{2}$,$\frac{1}{2}$) at the lowest temperature, T=66 mK and zero applied field.  Figures 4 b), c) and d) show the same elastic scattering over the same range in {\bf Q} space at T=66 mK and $\mu_0$H=0.25 T; T=1.5 K and $\mu_0$H=0 T; and T=60 K and $\mu_0$H=0 T; respectively. The diffuse elastic scattering shown in Fig. 4 a) is of the same ($\frac{1}{2}$,$\frac{1}{2}$,$\frac{1}{2}$) type that was measured on DCS, although it occurs here within a different scattering plane than was shown in Figs. 1, 2, and 3. Figure 4 e) shows cuts through this (-$\frac{1}{2}$,$\frac{1}{2}$,-$\frac{1}{2}$) position for different fields and temperatures, again using a T=66 mK and $\mu_0$H=7 T data set as a background. As seen in Fig. 4 e), the (-$\frac{1}{2}$,$\frac{1}{2}$,-$\frac{1}{2}$) elastic scattering can be completely removed at low temperatures by an applied field of 0.25 T along [11-1], confirming that it is magnetic and of the same origin as the elastic scattering shown in Figs. 1, 2, and 3. The elastic scattering also vanishes by T = 1.5 K, indicating that it is likely related to the broad hump in specific heat observed at 0.4 K in many samples \cite{GingrasPRB2000,Hamaguchi2004,YaouancPRBspecificheat,ChapuisPhD,Takatsu2012}. We note that earlier neutron measurements using DCS on this same single crystal sample did not observe this ($\frac{1}{2}$,$\frac{1}{2}$,$\frac{1}{2}$) Bragg-like magnetic scattering at T= 0.4 K and $\mu_0$H=0 T \cite{Kirrily2006}.

Figure 5 shows energy vs. wavevector slices of these same S({\bf Q}, E) data under the same conditions as were relevant for the elastic scattering maps in Fig. 4.  Figure 5 a) shows data at T = 66 mK and $\mu_0$H=0 T, Fig. 5 b) shows T=66 mK and $\mu_0$H=0.25 T, Fig. 5 c) shows T=1.5 K and $\mu_0$H=0 T, and Fig. 5 d) shows T=60 K and $\mu_0$H=0 T.  As with the elastic data shown in Fig. 4, this inelastic data has employed a background data set at T=66 mK and $\mu_0$H=7 T.  The data set at T=66 mK and zero field in Fig. 5 a) shows elastic scattering characteristic of the diffuse antiferromagnetic spin ice short range order.  This elastic scattering is separated from a gapped continuum of inelastic scattering which extends from $\sim$ 0.06 meV out to $\sim$ 0.4 meV.  Some magnetic inelastic spectral weight appears to exist within the gap, but this spectral weight is clearly strongly suppressed compared with that near the peak in the inelastic spectrum near $\sim$ 0.08 meV.  Application of a small field of 0.25 T at 66 mK, as is shown in Fig. 5 b), completely suppresses the elastic magnetic scattering at ($\frac{1}{2}$,$\frac{1}{2}$,$\frac{1}{2}$), as was shown in the elastic map in Fig. 4 b), but the low energy inelastic magnetic scattering below 0.2 meV is also strongly suppressed. Data at higher temperatures and zero field are shown in Fig. 5 c) for T=1.5 K and Fig. 5 d) for T=60 K. The inelastic scattering at 1.5 K is taken at sufficiently low temperature (below $|\Theta_{\textnormal{CW}}|$=$|$-19 K$|$) to still be within the cooperative paramagnetic regime. This 1.5 K data set shows the absence of ($\frac{1}{2}$,$\frac{1}{2}$,$\frac{1}{2}$) elastic magnetic scattering and an ungapped quasielastic magnetic spectrum. The data set at T=60 K and zero field in Fig. 5 d) is taken well into the fully paramagnetic phase. Its inelastic spectrum is also ungapped and quasielastic, and the overall bandwidth of the magnetic inelastic spectrum has softened by a factor of $\sim$ 1.5 relative to that shown in Fig. 5 a) for T=66 mK and zero field.

\begin{figure}[h]
\includegraphics[width=8cm]{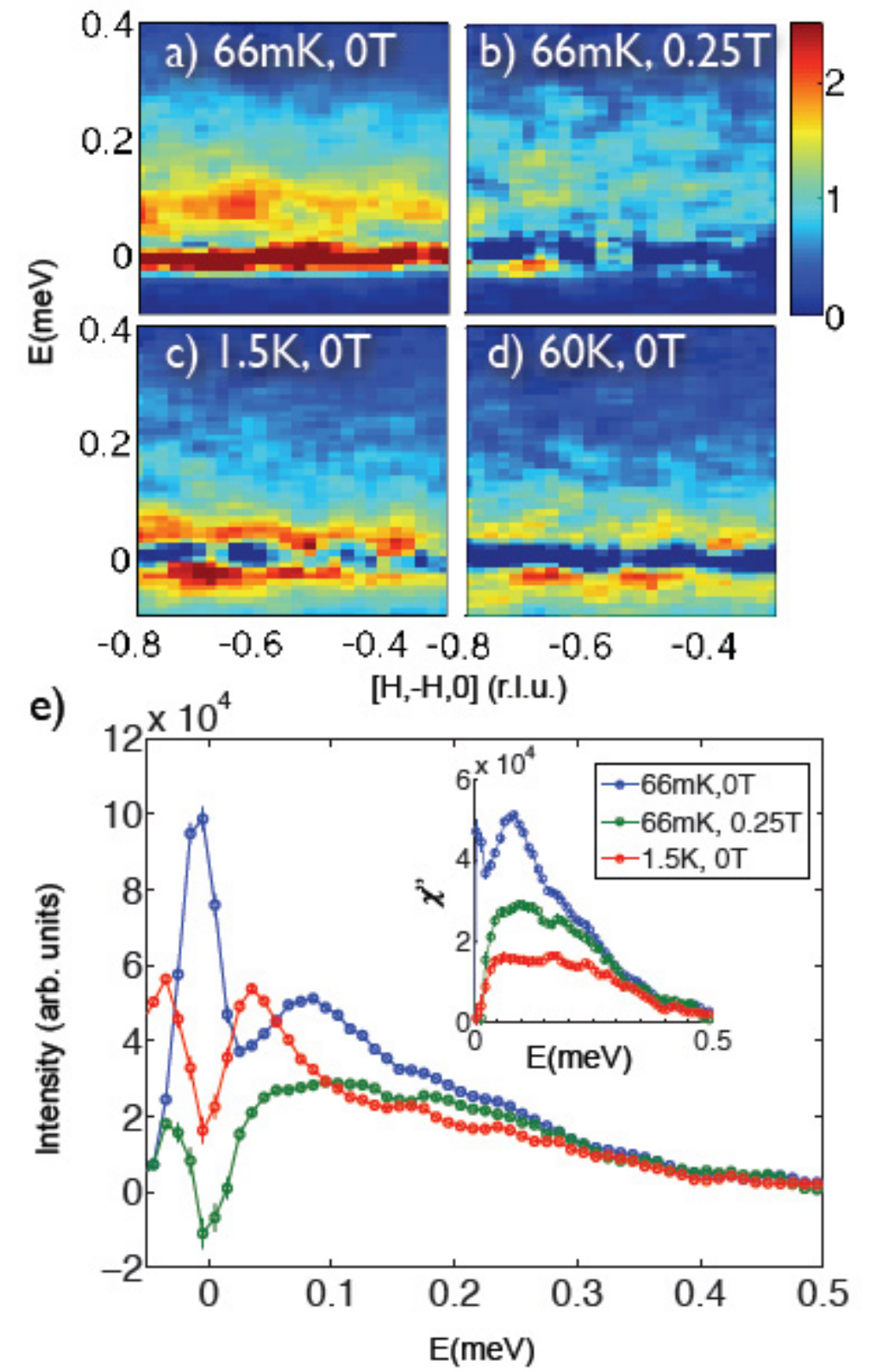}%
\caption{High energy resolution data from LET, with the 66 mK, 7 T data set used as a background. Inelastic scattering integrated over [$\frac{1}{2}$H,$\frac{1}{2}$H,H] = [-1.5,1.0] r.l.u. and [K,K,-K] = [-0.5, 0.5] r.l.u. is shown for a) T=66 mK and $\mu_0$H=0, b) T=66 mK and $\mu_0$H=0.25 T, c) T=1.5 K and $\mu_0$H=0, and d) T=60 K and $\mu_0$H=0.  A spin gap of $\sim$0.06-0.08 meV opens up below 1.5 K, which was not resolved previously. The spin gap correlates strongly with the appearance of the ($\frac{1}{2}$,$\frac{1}{2}$,$\frac{1}{2}$) elastic magnetic peaks shown in Figs. 1-4. e) Field and temperature dependence of the low energy inelastic scattering as shown in panels a) - c), integrated over the full [H,-H,0] range. The inset to Fig. 5 e) shows the same data corrected for the Bose factor, that is, $\chi''$({\bf Q}, E) (see Eq. 1).}
\label{figure}%
\end{figure}

The similarities and differences between the inelastic magnetic spectrum in the presence and absence of the ($\frac{1}{2}$,$\frac{1}{2}$,$\frac{1}{2}$) Bragg-like scattering can be appreciated by taking energy cuts through the data shown in Figs. 5 a) through 5 d). This is what is shown in Fig. 5 e) for T=66 mK and $\mu_0$H=0 T, for T=66 mK and $\mu_0$H=0.25 T, and for T=1.5 K and $\mu_0$H=0 T. One clearly sees that resolution-limited elastic magnetic scattering is only present for T=66 mK and zero field. The inelastic magnetic spectrum at this base temperature and zero field clearly peaks near 0.08 meV and is strongly suppressed at lower energies. The contrast between this gapped inelastic spectrum at 66mK and zero field and the quasielastic scattering at higher temperature at T=1.5K, but still low compared with $\Theta_{\textnormal{CW}}$ is clear from Fig. 5 e). The inset to Fig. 5 e) shows the dynamic susceptibility, $\chi''({\bf Q}, E)$, which is related to S({\bf Q}, E) through the Bose factor:
\begin{equation}
S({\bf Q},E)=\frac{\chi''({\bf Q},E)}{1-e^{-E/k_{B}T}}.
\end{equation}

We estimate the spin gap energy at $\sim$0.06-0.08 meV. The appearance of the gapped structure in the low energy magnetic excitation spectrum of Tb$_2$Ti$_2$O$_7$ is clearly tightly correlated with the appearance of strong ($\frac{1}{2}$,$\frac{1}{2}$,$\frac{1}{2}$) elastic magnetic Bragg scattering, which arises due to the formation of a SRO AF spin ice structure, as shown in Figs. 3 c) and d).

The clear presence of the $\sim$0.06-0.08 meV gap separating the low energy inelastic scattering (0.06-0.6 meV) from the elastic scattering due to the SRO AF spin ice scattering at ($\frac{1}{2}$,$\frac{1}{2}$,$\frac{1}{2}$) wavevectors confirms that three distinct regimes of magnetic scattering exist at low temperatures: elastic, low-energy inelastic, and ``high''-energy inelastic scattering due to the excited CF doublet at energies of $\sim$ 1 meV (see Fig. 1 d). True static long range order in zero field would be inconsistent with the non-magnetic, singlet ground state scenario previously suggested \cite{BonvilleSinglet2011}.

Pinch-point diffuse scattering characteristic of Coulombic spin ice correlations \cite{Morris2009,Bramwell2009,Fennell2009} has been of great interest in the classical spin ice materials Dy$_{2}$Ti$_{2}$O$_{7}$ and Ho$_{2}$Ti$_{2}$O$_{7}$ and recently in Tb$_{2}$Ti$_{2}$O$_{7}$ \cite{Fennell2012}. Figure 6 shows these correlations, most notably around (0,0,2), as they appear in the elastic scattering at T = 0.07 K and (a) zero field, and with a (b) 2 T magnetic field applied along [1-10].

\begin{figure}[h]
\includegraphics[width=8cm]{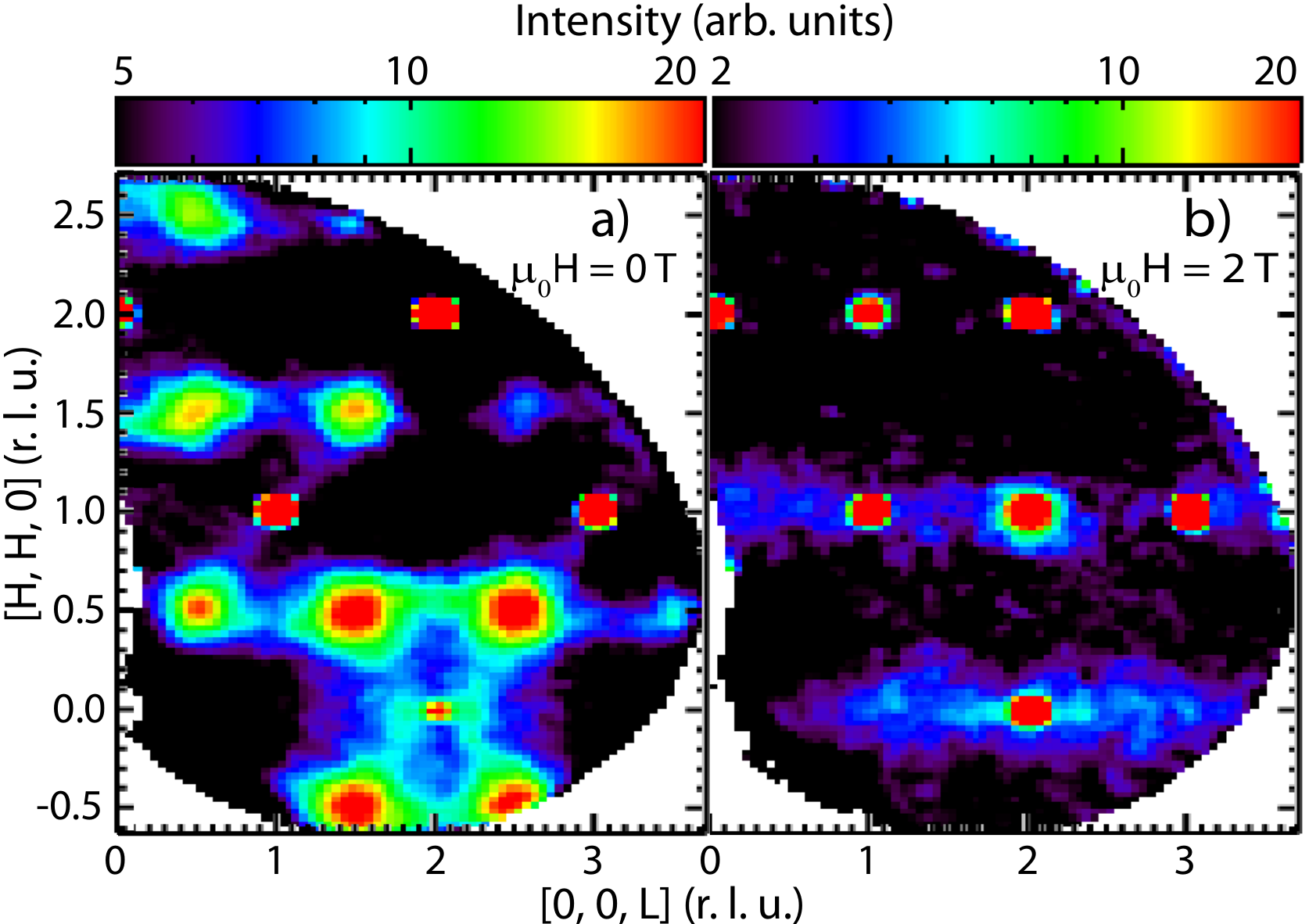}%
\caption{Pinch-point like scattering in the ground state of Tb$_{2}$Ti$_{2}$O$_{7}$ observed on DCS at $\sim$ 70 mK for $\mu_0$H = 0 T in panel a). Panel b) shows diffuse scattering along the $<$0,0,L$>$ direction under application of a $\mu_0$H = 2 T field, which might originate from anisotropic exchange. Both data sets show elastic scattering for -0.1 meV $<$ E $<$ 0.1 meV and are displayed on a logarithmic intensity scale (compared to a linear scale in Fig. 1).}%
\label{figure6}%
\end{figure}

Our pinch-point diffuse scattering near (0,0,2) is qualitatively similar to that observed in Fig. 2a) of Fennell {\it et al.} \cite{Fennell2012}. As shown in Fig. 6, we observe ``rods'' of scattering along the $<$0,0,L$>$ and $<$H,H,H$>$ directions, which appear to be modulated by pinch points near the zone centers at (1,1,1), (1,1,3) and near (0,0,2). In a 2 T field applied along [1-10], all elastic scattering at ($\frac{1}{2}$,$\frac{1}{2}$,$\frac{1}{2}$) positions has disappeared and only broad diffuse scattering along the $<$0,0,L$>$ direction remains. The origin of this rodlike elastic diffuse scattering is not understood, but one can speculate that it originates from anisotropic exchange in Tb$_{2}$Ti$_{2}$O$_{7}$, as is known to describe the microscopic spin Hamiltonian in Yb$_{2}$Ti$_{2}$O$_{7}$ \cite{KatePRX} and Er$_{2}$Ti$_{2}$O$_{7}$ \cite{Savary}.

\section{Conclusion}

The ground state of Tb$_{2}$Ti$_{2}$O$_{7}$ has by now been studied for more than a decade by neutron scattering techniques. It is surprising that the ($\frac{1}{2}$,$\frac{1}{2}$,$\frac{1}{2}$) elastic SRO scattering, presaging a transition to LRO, has not previously been observed. This is likely due to the low onset temperature of this scattering: for example Rule {\it et al.} \cite{Kirrily2006} previously studied the same single crystal with the same instrumental conditions at DCS, but at temperatures of 400 mK and above, without seeing this scattering. In fact, triple-axis measurements by Yasui {\it et al.} \cite{Yasui2002} do show evidence for a peak at ($\frac{1}{2}$,$\frac{1}{2}$,$\frac{5}{2}$) at T = 0.4 K (Fig. 2 in Ref. 44), however this feature was not pursued further in subsequent studies. Diffraction experiments, such as that by Gardner {\it et al.} \cite{Jason2001}, would be problematic due to integration over elastic and low-energy inelastic scattering, which we show in Fig. 1 to have very different {\bf Q} dependencies.  

In conclusion, new neutron scattering measurements on Tb$_{2}$Ti$_{2}$O$_{7}$ at T = 0.07 K in zero field have revealed elastic diffuse scattering at ($\frac{1}{2}$,$\frac{1}{2}$,$\frac{1}{2}$) positions characteristic of short-range AF spin ice correlations extending over roughly two conventional pyrochlore unit cells. This elastic scattering is separated by a gap of $\sim$0.06-0.08 meV from low lying inelastic scattering, and can be quantitatively described based on an ordered two-in, two-out local spin ice structure with a spin canting angle of $\sim$ 12$^\circ$. This development of AF spin ice correlations in Tb$_{2}$Ti$_{2}$O$_{7}$ is characteristic of a strong tendency to form the corresponding LRO state which does not occur either because the temperature is not sufficiently low, or due to weak disorder in the samples, as has been recently characterized in other exotic pyrochlore magnets \cite{KateYbTiOstructure}, or both.

Following completion of this work, we became aware of two preprints reporting related neutron work on single crystal \cite{PetitarXiv2012} and polycrystalline \cite{KadowakiTTOarXiv} Tb$_{2}$Ti$_{2}$O$_{7}$ samples.

\section*{Acknowledgments}
The authors acknowledge useful contributions from M. J. P. Gingras. This work utilized facilities supported in part by the National Science Foundation under Agreement No. DMR-0944772, and was supported by NSERC of Canada. The DAVE software package \cite{DAVE} was used for data reduction and analysis of DCS data. LET data were reduced using Mantid \cite{mantidproject} and analysed using the HORACE software package \cite{Horace}.

\end{document}